\documentclass[aps,pre,twocolumn,groupedaddress]{revtex4-2}
\usepackage{graphicx}
\usepackage{dcolumn}
\usepackage{bm}
\begin{document}


\title{Is the atomic metal vapor a dielectric state?}


\author{Alexander L. Khomkin}
\affiliation{Joint Institute for High Temperatures of RAS, Izhorskaya 13 Bldg. 2, 125412, Moscow, Russia} 
\author{Aleksey S. Shumikhin}
\email{shum$_$ac@mail.ru}
\affiliation{Joint Institute for High Temperatures of RAS, Izhorskaya 13 Bldg. 2, 125412, Moscow, Russia}


\date{\today}

\begin{abstract}
 We propose a simple method for calculating the metal vapor conductivity at the critical point and near-critical isotherms. This method's base is the hypothesis of an electron jellium’s existence as an origin of the conduction band in metal vapor's gaseous phase. Satisfactory agreement with the experimental data for alkali metals (Cs, Rb) concludes that the ``cold ionization" mechanism is possible in the critical point's vicinity. The liquid metal--dielectric transition, as previously thought, does not occur; but the liquid metal--gaseous metal transition occurs probably. The gaseous metal is not a dielectric state of the metal.
\end{abstract}


\maketitle

\section{Introduction}
Generally, atomic gases are considered as a dielectric state of a substance that does not conduct an electric current at low temperatures until thermal ionization begins. However, under isothermal compression, atomic metal vapor can pass into a liquid state and then into a solid-state with high, metallic conductivity. This transition occurs with a jump in density at temperatures below critical. We can talk about the coexistence of two transitions: vapor--liquid and dielectric--metal~\cite{ZL}. However, analyzing the possible topology of these transition locations on the phase plane, the authors~\cite{ZL} mentioned, as a hypothetical, the possibility of metal's existence in the gaseous state. 

The transition to the conduction state occurs exponentially, without jumps, at higher temperatures (near-critical) and we can talk about the process of metallization~\cite{KhomkinShumikhin2017} --- the origin of metal properties. 

The transformation of the solid metal conductivity band occurs in the reverse process --- the expansion process. In a specific form, it continues to be present in the liquid phase. Electrons from the conduction band begin, fully or partially, to gradually return to the bound orbits of atoms in the complex expansion process. Embedded Atom Model (EAM)~\cite{Daw,Puska,Bel2013} used to describe liquid metals regards atoms as particles immersed in an electron jellium, a transformed conduction band. The jellium density is the sum of the contributions of the electron density of the atom bound states of the first coordination sphere. Most often, density calculating performs by using semi-phenomenological schemes. 

These two processes (gas compression and liquid metal expansion) converge at a critical point. We assumed~\cite{KhomkinShumikhin2016} that the conduction band's traces in the jellium's form will remain in the gas phase. The gaseous phase's jellium consists of the wave function tails of bound electrons lying outside the Wigner-Seitz cell (WS). The possibility of conducting jellium's existence in the gaseous phase is indicated by the experimentally measured and sufficiently high electrical conductivity of alkali metal vapor (Cs, Rb) at the critical point and in the near-critical region at low temperatures ($T\sim 2000$~K)~\cite{Renkert,Hensel1979,Hensel1980}.

\section{Liquid branch of binodal. The electrical conductivity}

The temperature dependencies of the electrical conductivity and density of Cs and Rb were measured by Hensel with co-authors~\cite{Renkert,Hensel1979,Hensel1980} on the binodal's liquid branch from the melting point up to the critical point. Knowing the density and the temperature at binodal, it is possible to calculate the conductivity and compare it with experimental data. Using modern modifications of the Ziman theory~\cite{Ziman} for metals, it is possible to successfully describe the conductivity on the liquid branch of the binodal (see, for example, \cite{Redmer1992,Apfelbaum}) up to the critical point. However, upon reaching a precisely critical point, calculations give significantly higher conductivity values than those measured experimentally. The authors \cite{Redmer1992,Apfelbaum} quite rightly noted that the process of conduction electrons returning to the bound state with the ion core begins at approaching the critical point. The Ziman theory does not take this process into account.

\section{Gaseous branch of binodal. Thermal ionization}

The approach to the critical point performs by moving along the gas branch of the binodal. The early theoretical works associate the exponential increase in the conductivity of alkali metal vapor during compression with a well-known effect in plasma physics --- an increase in the number of thermally ionized electrons caused by the ionization potential lowering of the atom in the ionization equilibrium equation --- the Saha formula. Works \cite{Vedenov,Gryko} consider contributing to the ionization potential lowering from the electron-atom and the ion-atom interactions. The linear dependence of corrections on the atom density in the ionization potential lowering leads to an exponential increase of the free electron density in weakly ionized vapor, but it was not enough. Further development of the theory followed the path of searching for effects that lead to more significant ionization potential lowering. Let us note a series of works on cluster-drop models \cite{Likalter1978,Yakubov,Khrapak,Zhuch}. The assumption was that the conversion of ions to cluster \cite{Likalter1978,Khrapak} or drop \cite{Yakubov,Zhuch} ions would lead to a more significant increase of the free electrons concentration. However, the obtained increase in the electron concentration was insufficient. One should note that many parameters of small-sized clusters taken into account in the models \cite{Likalter1978,Khrapak} were unknown at that time. 

Numerical calculations of small clusters properties (neutral, positively, and negatively charged) for alkali metals \cite{Ali} allowed to perform piece-by-piece calculating \cite{KhomkinShumikhin2013}. These calculations showed that the conversion of ions does occur, simultaneously both for positive and negative ions. The neutral component of the plasma of alkali metal vapor near the binodal was predominantly atomic with a small admixture of molecules (it confirmed further by other independent calculations \cite{Semenov}). The concentration of the charged component was small. Contrary to the predictions [17-20], it turned out to be predominantly ionic in composition \cite{KhomkinShumikhin2013}, which in cluster models led naturally to relatively low conductivity values of near-critical alkali metal vapor. 

Significant progress in the metallization processes understanding for alkali metal vapor occurred after the appearance of a series of works by A.\,A. Likalter (see reviews \cite{Likalter1992,Likalter2000}). Although he was the author of the cluster model of ion conversion \cite{Likalter1978}, he refused to find a ``key" ionization potential lowering. Likalter suggested that in the near-critical region, the conductivity's crucial role will not be played by thermally ionized, free electrons, but by bound electrons, whose classically available orbitals begin to overlap. Overlapping orbitals leads to the formation of ``percolation" clusters, through which the current passes. One can talk about the appearance of the conduction band's origin in the gaseous phase. The ideas of A.A. Likalter were very constructive and allowed to calculate the conductivity at near-critical isotherms, but not at critical points. An equation of state, which allowed to estimate the parameters of the critical points for alkali and many other metals, was proposed for ``quasi-atoms", atoms with overlapping classical orbits, in work \cite{Likalter1997}. 

In our opinion, the works of Likalter \cite{Likalter1992,Likalter2000} put an end to attempts to explain the conductivity of alkali metal vapor in the vicinity of the critical point by the conductivity of thermally ionized electrons and further searching for a ``key" ionization potential lowering. 

We should note another essential aspect of the metallization problem, which formed the basis for our idea and method of calculating the gaseous metal vapor state's jellium concentration. It is related to the necessity of the correction of the concept of an ``isolated atom". An atom is a particle consisting of a nucleus and bound electrons. The atom has an extended internal structure. The Schr{\"o}dinger equation is solved to calculate the spectrum and wave functions of the atom bound states with the conventional boundary condition: the wave functions tend to zero ``at infinity". Above is the ``isolated atom" model. The spectrum of bound states is used, for example, for calculation of the partition function of an atom. For calculating the equation of state, composition, and transport coefficients, the obtained wave functions of bound electrons of an atom are present in an implicit form. They are non-zero in the entire space, ``ad infinitum". Since gas with the atom density $n_a$, the volume per atom is always limited; then, it is necessary to correct the ``isolated atom" approximations at approaching the critical region. The volume per atom is determined by the WS cell volume with radius $R_a$: 
\begin{equation}
  R_a = \left(\frac{3}{4\pi n_{a}}\right)^{1/3}. 
\label{eq1}
\end{equation} 

The wave function of a bound electron decreases exponentially at infinity, and therefore the effect of the limiting volume per atom in a rarefied gas is small. We draw attention to the fact that when approaching the critical point of metal vapor, the fraction of the bound electrons density that lies outside the WS cell turns out to be quite significant. These electrons in our model form an electron jellium. The contribution to the jellium is given by the tails of the electron density of bound electrons from all atoms of the system. The fraction of the jellium electrons density $n_{\text j}$ from the density of atoms $n_a$ can be called the degree of ``cold ionization" $\alpha_{\text j}=n_{\text j}/n_{a}$. 

\section{Calculation of $\alpha_{\text j}$ in the cell approximation using Hartree-Fock-Slater wave functions}

As a first approximation of the solution of a very complex problem about the distribution of electron density in an atomic gas, let us consider the following, in our opinion, a quite reasonable procedure. 

Suppose we know the bound i-th electron $\Psi^{(\text i)}(r)$ wave function in the ``isolated atom" approximation. Contribution to jellium comes from wave function tails of bound electrons of all atoms: both from this and the surrounding ones. It is possible to calculate in the first approximation the fraction of the electron density involved in the formation of jellium in the cell approximation. The $\alpha_{\text j}^{(\text i)}$ value is determined by integrating $\Psi^{(\text i)}(r)^2$ outside the Wigner-Seitz cell (this corresponds to the part from surrounding cells to this one) and the permanent background within the cell $\Psi^{(\text i)}(y_a)^2$ (this corresponds to the contribution to the jellium from this cell): 
\begin{equation}
  \alpha_{\text j}^{(\text i)} = \int_{y_a}^{\infty}{\Psi^{(\text i)}(r)^2 r^2 dr} + \frac{y_{a}^{3}}{3}\Psi^{(\text i)}(y_a)^2, 
\label{eq2}
\end{equation} 
where $y_a=R_{a}/a_{0}$ is the radius of the atomic Wigner-Seitz cell in atomic units. The total electron density is conserved and the Wigner-Seitz cell is electroneutral. 

Work \cite{Clementi1974} presents the wave functions of an isolated atom calculated numerically by the Hartree-Fock method. The data covers all elements up to the atomic number $Z = 54$. In \cite{Sachdeva,Clementi1967}, you can also find data for heavier elements. 

The wave function $\Psi^{(\text i)}(r)$ of an arbitrary i-th atomic electron in a specific quantum state presented below in the form of expansion of the Slater-type orbitals $\chi_{\lambda p}(r,\theta,\varphi)$: 
\begin{equation}
  \Psi^{(\text i)}(r) = \sum_{\lambda, p} C_{\lambda, p}\chi_{\lambda, p}(r,\theta,\varphi). 
\label{eq3}
\end{equation}

The coefficients for Slater-type orbitals (3) are presented in \cite{Clementi1974,Sachdeva} as tables for each electron states. 
Formally, we can calculate the $\alpha_{\text j}^{(\text i)}$ values for all the electrons of an arbitrary atom. Their sum will give an estimate of the desired degree of ``cold ionization". In our calculations, we used data \cite{Clementi1974,Sachdeva,Clementi1967} only for valence electrons, since the contribution of the ion core electrons in our conditions is small and does not affect the final value of $\alpha_{\text j}=\sum_{\text i}{\alpha_{\text j}^{(\text i)}}$. Moreover, one should remember that in the vicinity of the critical point, even valence electrons participate in the formation of the jellium only partially. When approaching the metal’s normal density, all valence electrons are involved in the jellium formation, and $\alpha_{\text j}$ tends to full valence. 

\begin{figure}[b]
\includegraphics{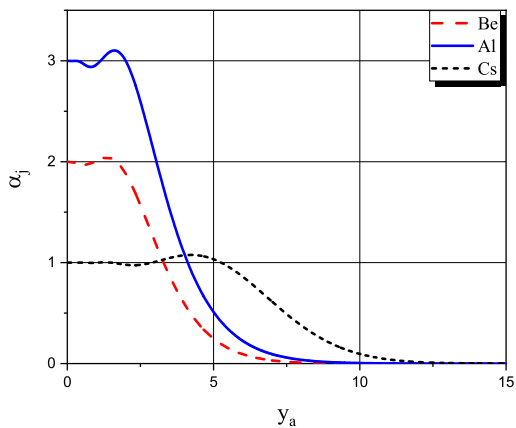}
\caption{\label{fig1} The degree of ``cold ionization" for Be, Al, Cs depending on $y_a$ --- radius of the Wigner-Seitz cell in atomic units.}
\end{figure}

Figure~\ref{fig1} shows our calculations of the degree of ``cold ionization" $\alpha_{\text j}$ for various metals, depending on $y_a$ by the ratio~(\ref{eq2}). It is small at low densities (large $y_a$). The value $\alpha_{\text j}$ tends to the valence of the element with density increase. For example, at the critical point of Al, $y_a \sim 5$, and for the metal in the normal state $y_a \sim 3$. The critical density of Cs corresponds to $y_a \sim 9.8$.

The calculated dependence's analysis of ``cold ionization" degree $\alpha_{\text j}$ on the density ($y_a$) allows us to conclude that the electron’s state in dense metal vapor is no longer purely bound, as previously thought, but rather mixed, as noted by Likalter~\cite{Likalter1992}. An electron with negative energy simultaneously resides in a bound localized state and a delocalized state of jellium. 

\section{Calculation of the electrical conductivity. Results and Discussion} 

Jellium electrons can move from cell to cell. It is natural to assume that this will mainly be the movement between neighboring cells compared to the metal's conduction electrons. As a result, conductivity appears in a ``cold" atomic gas without thermal ionization processes. To estimate jellium electrons' conductivity, we use the Regel-Ioffe formula for minimal metallic conductivity \cite{RegelIoffe}, which considers the transfer mechanism between cells: 
\begin{equation}
  \sigma = n_{\text j}\frac{q_{e}^{2}}{m_{e}}\tau,  
\label{eq4}
\end{equation} 
where $q_e$ is the charge, and $m_e$ is the mass of an electron; $n_{\text j}=\alpha_{\text j} n_{a}$ is the jellium density; $\tau$ is the mean free time. The mean free time is equal to the transit flight time of inter-nuclear distance (twice the radius of the Wigner-Seitz cell in atomic units) with Fermi velocity $v_{\text F}=p_{\text F}/m_e$: 
\begin{equation}
  \frac{\tau}{m_e} = \frac{2R_a}{p_{\text F}},  
\label{eq5}
\end{equation} 
where $p_{\text F}=(3\pi^{2}n_{\text j})^{1/3}$ is the Fermi momentum. 
\begin{equation}
  \sigma_{\text j} = n_{\text j}^{2/3}\frac{q_{e}^{2}}{9\cdot 10^{11}}\frac{2y_{a}a_{0}}{(3\pi^2)^{1/3}\hbar}.
\label{eq6}
\end{equation}

We can determine the jellium electrons’ conductivity by their concentration $n_{\text j}$, related to the density of atoms, and a direct relationship to the density of atoms via $y_a$ --- the atomic cell's size in atomic units. Temperature dependence is absent. The dimension of all values in (\ref{eq6}) is CGSE, and the conductivity’s one is in 1/($\Omega\cdot$ cm). To estimate the vapor conductivity for Cs, Al, and Be, it is sufficient to set their density, find $\alpha_{\text j}$ from the graphs shown in figure \ref{fig1}, and use (\ref{eq6}) to calculate the conductivity.
\begin{equation}
  \rho \rightarrow n_{a} \rightarrow y_{a} \rightarrow n_{\text j}=\alpha_{\text j} n_{a} \rightarrow \sigma. 
\label{eq7}
\end{equation} 

Table~\ref{table1} shows a step-by-step calculation of some metals' conductivity at critical points (Be, Al, Rb, Cs). The experimental density at the critical point indicated for cesium and rubidium \cite{Hensel1985}. The critical density, calculated in \cite{KhomkinShumikhin20171}, showed for aluminum and beryllium. We can calculate the degree of ``cold ionization" $\alpha_{\text j}$, but we can also estimate it from Fig.~\ref{fig1}. Experimentally measured conductivity values for cesium and rubidium at the critical point are of the order of $250\pm 150$~1/($\Omega\cdot$ cm)~\cite{Hensel1980}. The conductivity value at the critical point is not known for other metals.

\begin{table}[b]
\caption{\label{table1}
Step-by-step calculation of some metals' conductivity at critical points (Be, Al, Rb, Cs).}
\begin{ruledtabular}
\begin{tabular}{ccccccc}
 Metal&$\rho_{\text c}$ (g/cm$^3$)&$n_a$ (cm$^{-3}$)&$y_a$&$\alpha_{\text j}$ &$n_{\text j}$ (cm$^{-3}$)&$\sigma$ (1/$\Omega\cdot$cm) \\
 \hline
Cs& 0.38 & 1.7$\cdot 10^{21}$ & 9.79 & 0.12 & 2.4$\cdot 10^{20}$ & 272 \\
Rb& 0.29 & 2.04$\cdot 10^{21}$ & 9.23 & 0.18 & 3.67$\cdot 10^{20}$ & 394 \\
Be& 0.38 & 2.53$\cdot 10^{22}$ & 4.0 & 0.6 & 1.51$\cdot 10^{22}$ & 2000 \\
Al& 0.6 & 1.33$\cdot 10^{22}$ & 4.94 & 0.54 & 7.15$\cdot 10^{21}$ & 1970 \\
\end{tabular}
\end{ruledtabular}
\end{table}

\begin{figure}[b]
\includegraphics{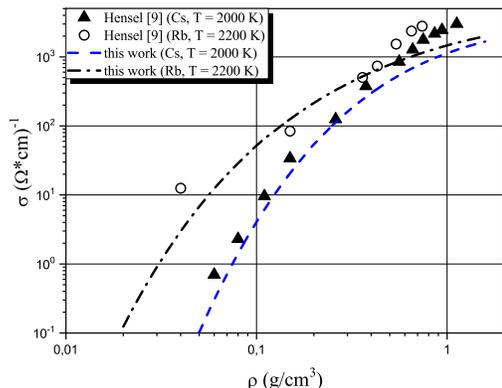}
\caption{\label{fig2} The electrical conductivity of alkali metal vapor for near-critical isotherms vs. density. Experiments: filled triangles and open circles correspond to Cs and Rb data from \cite{Hensel1980}. Theory: dashed and dash-dotted curves --- our calculations using (\ref{eq6}) for Cs and Rb.}
\end{figure}

Figure~\ref{fig2} shows experimental data \cite{Hensel1980} for Cs and Rb vapor's electrical conductivity on near-critical isotherms and our calculations using the formula (\ref{eq6}). The degree of ``cold ionization" was calculated using (\ref{eq2}). 

The comparison with the experimental data for the conductivity of alkali metal vapor at the critical point and its vicinity allows us to conclude the emergence of the jellium --- the origin of the conduction band in the gaseous phase. The emergence of the jellium leads to a new effect --- the process of ``cold ionization" and a new conduction channel. A particular discrepancy between our calculations and the experiment at densities $\rho\sim 1$~g/cm$^3$ is due to the long-range order's origin and the structure's appearance. We plan to take this effect into account in future works. 

\section{Conclusions} 

Metal vapor in the vicinity of the critical point is not an absolute dielectric due to the jellium's existence --- the conduction band’s origin. Our calculations of metal vapor conductivity at the critical point and its vicinity quite convincingly confirmed this hypothesis. It is more accurate to speak about the existence of a transition: a liquid metal --- a gaseous metal instead of the metal-dielectric transition and the process of ``cold" metallization at compression.

\begin{acknowledgments}
The work is supported by the grant in the form of a subsidy for a large scientific project in priority areas of scientific and technological development No.~13.1902.21.0035.
\end{acknowledgments}

\bibliography{apssamp_khomkin}

\end{document}